\documentclass[12pt]{article}
\usepackage{amsmath}

\begin{document}

\baselineskip=17pt

\begin{titlepage}
% \rightline{\today}
\rightline{\tt arXiv:0704.3612}
\rightline{\tt DESY 07-056}
\begin{center}
\vskip 3.0cm
{\Large \bf {Real analytic solutions for marginal deformations}}\\
\vskip 0.4cm
{\Large \bf {in open superstring field theory}}
\vskip 1.0cm
{\large {Yuji Okawa}}
\vskip 1.0cm
{\it {DESY Theory Group}}\\
{\it {Notkestrasse 85}}\\
{\it {22607 Hamburg, Germany}}\\
yuji.okawa@desy.de

\vskip 3.0cm

{\bf Abstract}
\end{center}

\noindent
We construct analytic solutions for marginal deformations
satisfying the reality condition
in open superstring field theory formulated by Berkovits
when operator products
made of the marginal operator
and the associated superconformal primary field
are regular.
Our strategy is based on the recent observation by Erler
that the problem of finding solutions
for marginal deformations in open superstring field theory
can be reduced to a problem in the bosonic theory
of finding a finite gauge parameter
for a certain pure-gauge configuration
labeled by the parameter of the marginal deformation.
We find a gauge transformation
generated by a real gauge parameter
which infinitesimally changes the deformation parameter
and construct a finite gauge parameter
by its path-ordered exponential.
The resulting solution satisfies
the reality condition by construction.

\end{titlepage}

\newpage

% \tableofcontents

\section{Introduction}
\setcounter{equation}{0}

Analytic methods
in open bosonic string field theory
\cite{Witten:1985cc}\footnote{
See \cite{Taylor:2003gn, Sen:2004nf, Rastelli:2005mz,
Taylor:2006ye} for reviews on string field theory.}
triggered by Schnabl's construction of an analytic solution
for tachyon condensation \cite{Schnabl:2005gv}
and further developed
in \cite{Okawa:2006vm, Fuchs:2006hw, Fuchs:2006an, Rastelli:2006ap,
Ellwood:2006ba, Fuji:2006me, Fuchs:2006gs, Okawa:2006sn,
Asano:2006hk, Asano:2006hm, Erler:2006hw, Imbimbo:2006tz,
Erler:2006ww, Schnabl:2007az, Kiermaier:2007ba}
have recently been extended
to open superstring field theory formulated
by Berkovits \cite{Berkovits:1995ab},
and analytic solutions for marginal deformations
were constructed in \cite{Erler:2007rh, Okawa:2007ri}.\footnote{
For earlier study of marginal deformations in string field theory
and related work,
see \cite{Sen:2000hx, Iqbal:2000qg, Takahashi:2001pp, Marino:2001ny,
Kluson:2002ex, Takahashi:2002ez, Kluson:2002hr, Kluson:2002av,
Kluson:2003xu, Coletti:2003ai, Berkovits:2003ny, Sen:2004cq,
Katsumata:2004cc, Yang:2005iu, Kishimoto:2005bs}.
}
The solutions are surprisingly simple
and very similar to those in open bosonic string field theory
constructed in \cite{Schnabl:2007az, Kiermaier:2007ba}.
However, the reality condition on the open superstring field
was not satisfied.
While we expect that the solution
in \cite{Erler:2007rh, Okawa:2007ri}
is equivalent to a real one by a gauge transformation,
it is desirable to find an analytic solution
satisfying the reality condition.
In this paper we explicitly construct
a real analytic solution.

The equation of motion in open superstring field theory
\cite{Berkovits:1995ab} is
\begin{equation}
\eta_0 \, ( \, e^{-\Phi} \, Q_B \, e^\Phi \, ) = 0 \,,
\label{equation-of-motion}
\end{equation}
where $\Phi$ is the open superstring field
and $Q_B$ is the BRST operator.
The superghost sector is described
by $\eta$, $\xi$, and $\phi$
\cite{Friedan:1985ge, Polchinski:1998rr},
and $\eta_0$ is the zero mode of $\eta$.
All the string products in this paper
are defined by the star product introduced in \cite{Witten:1985cc}.
For any marginal deformation
of the boundary conformal field theory (CFT)
for the open superstring,
there is an associated superconformal primary field $V_{1/2}$
of dimension~$1/2$,
and the marginal operator $V_1$ of dimension~$1$
is the supersymmetry transformation of $V_{1/2}$.
In open superstring field theory \cite{Berkovits:1995ab},
the solution to the linearized equation of motion
associated with the marginal deformation
is given by the Grassmann-even state $X$
corresponding to the operator
$V (0) = c \, \xi e^{-\phi} V_{1/2} (0)$
in the state-operator mapping.
When the deformation is exactly marginal,
we expect a solution to (\ref{equation-of-motion})
of the following form:
\begin{equation}
\Phi_\lambda = \sum_{n=1}^\infty \lambda^n \, \Phi^{(n)}
\end{equation}
with $\Phi^{(1)} = X$,
where $\lambda$ is the deformation parameter.
The goal of the paper is to construct $\Phi^{(n)}$
satisfying the reality condition
when operator products made of $V_1$ and $V_{1/2}$ are regular.

In \cite{Erler:2007rh} Erler proposed
to solve the following equation:
\begin{equation}
e^{-\Phi} \, Q_B \, e^\Phi = \Psi_\lambda \,,
\label{Erler-equation}
\end{equation}
where $\Psi_\lambda$ is obtained from the solution
for marginal deformations in open bosonic string field theory
constructed in \cite{Schnabl:2007az, Kiermaier:2007ba} by replacing
the state corresponding to $cV_b (0)$ for the bosonic string
with the state $Q_B X$ for the superstring,
where $V_b$ is the marginal operator in the bosonic theory.
The state $\Psi_\lambda$ satisfies the equation of motion
in open bosonic string field theory,
\begin{equation}
Q_B \, \Psi_\lambda + \Psi_\lambda^2 = 0 \,,
\end{equation}
and to linear order in $\lambda$ it reduces to
\begin{equation}
\Psi_\lambda = \lambda \, Q_B X + O(\lambda^2) \,.
\end{equation}
Thus $\Psi_\lambda$ is a pure-gauge solution
generated by $Q_B X$,
and we expect a solution to (\ref{Erler-equation})
of the form
\begin{equation}
\Phi = \lambda \, X + O(\lambda^2) \,.
\end{equation}
Furthermore, $\Psi_\lambda$ is annihilated by $\eta_0$
because the state $X$ satisfies the linearized equation of motion
$\eta_0 Q_B X = 0$.
Therefore, the solution to (\ref{Erler-equation})
solves the equation of motion in open superstring field theory
(\ref{equation-of-motion}),
and the problem of solving the superstring theory
has been reduced to a problem in the bosonic theory.
A simple solution to (\ref{Erler-equation})
was obtained in \cite{Erler:2007rh},
but the reality condition
on the open superstring field was not satisfied.
The same solution was also obtained in \cite{Okawa:2007ri}
by a different approach.

Let us now consider the equation obtained from
(\ref{Erler-equation}) by taking a derivative
with respect to $\lambda$.
Since the left-hand side of (\ref{Erler-equation})
takes the form of a pure-gauge configuration
with respect to the gauge transformation in the bosonic theory,
its infinitesimal change must be written
as an infinitesimal gauge transformation
generated by some gauge parameter which we call $G(\lambda)$:
\begin{equation}
Q_B \, G (\lambda) + [ \, \Psi_\lambda , G (\lambda) \, ]
= \Psi'_\lambda \,, \qquad
\Psi'_\lambda \equiv \frac{d}{d \lambda} \, \Psi_\lambda \,.
\label{G-equation}
\end{equation}
Then a solution to (\ref{Erler-equation}) can be constructed
by a path-ordered exponential of $G(\lambda)$ as
\begin{equation}
e^{\Phi_\lambda} = {\rm Pexp} \, \biggl[ \,
\int_0^\lambda d \lambda' \, G(\lambda') \, \biggr] \,,
\end{equation}
or
\begin{equation}
\Phi_\lambda = \ln {\rm Pexp} \, \biggl[ \,
\int_0^\lambda d \lambda' \, G(\lambda') \, \biggr] \,.
\end{equation}
If $G(\lambda)$ satisfies the reality condition,
the solution $\Phi_\lambda$ also satisfies the reality condition
by construction.
This is our strategy for constructing a real solution
in open superstring field theory.
It turns out that it is easy to find a real solution
to (\ref{G-equation}).

\bigskip

After we completed the construction
of solutions satisfying the reality condition,
we learned that T.~Erler independently constructed
analytic solutions satisfying the reality condition
by a different approach.
His solutions were presented
in the second version of \cite{Erler:2007rh}.

\section{Pure-gauge string field}
\setcounter{equation}{0}

Let us begin with describing $\Psi_\lambda$
in (\ref{Erler-equation}).
It is obtained from the solution for marginal deformations
in open bosonic string field theory
constructed in \cite{Schnabl:2007az, Kiermaier:2007ba}
by replacing $cV_b$ in the bosonic theory
with the BRST transformation
of $V = c \, \xi e^{-\phi} V_{1/2}$
for the superstring.
This section largely overlaps with section 2 of \cite{Okawa:2007ri},
where the solution in open bosonic string field theory
was reviewed.
The string field $\Psi_\lambda$ is defined
by an expansion with respect to $\lambda$ as follows:
\begin{equation}
\Psi_\lambda = \sum_{n=1}^\infty \lambda^n \, \Psi^{(n)} \,.
\end{equation}
The BPZ inner product
$\langle \, \varphi, \Psi^{(n)} \, \rangle$
with a state $\varphi$ in the Fock space is given by
\begin{equation}
\begin{split}
\langle \, \varphi, \Psi^{(n)} \, \rangle
& = \int_0^1 dt_1 \int_0^1 dt_2 \ldots \int_0^1 dt_{n-1} \,
\langle \, f \circ \varphi (0) \, U (1) \, {\cal B} \,
U (1+t_1) \, {\cal B} \, U (1+t_1+t_2) \, \ldots \\
& \qquad \qquad \qquad \qquad \qquad \qquad \qquad {}\times
{\cal B} \, U (1+t_1+t_2+ \ldots +t_{n-1}) \,
\rangle_{{\cal W}_{1+t_1+t_2+ \ldots +t_{n-1}}} \,,
\end{split}
\label{Psi^(n)}
\end{equation}
where $U$ is the BRST transformation of $V$:
\begin{equation}
U (z) = Q_B \cdot V (z) \,, \qquad
V (z) = c \, \xi e^{-\phi} V_{1/2} (z) \,.
\end{equation}
We follow the notation used in \cite{Okawa:2006vm, Okawa:2006sn,
Kiermaier:2007ba}. In particular, see the beginning of section 2
of \cite{Okawa:2006vm} for the relation to the notation
used in \cite{Schnabl:2005gv}. 
Here and in what follows we use $\varphi$ to denote
a generic state in the Fock space
and $\varphi (0)$ to denote its corresponding operator
in the state-operator mapping.
We use the doubling trick
in calculating CFT correlation functions.
As in \cite{Okawa:2006sn},
we define the oriented straight lines $V^\pm_\alpha$ by
\begin{equation}
\begin{split}
& V^\pm_\alpha = \Bigl\{ \, z \, \Big| \,
{\rm Re} (z) = {}\pm \frac{1}{2} \, (1+\alpha) \, \Bigr\} \,, \\
& \hbox{orientation} \, :
{}\pm \frac{1}{2} \, (1+\alpha) -i \, \infty
\to {}\pm \frac{1}{2} \, (1+\alpha) +i \, \infty \,,
\end{split}
\end{equation}
and the surface ${\cal W}_\alpha$ can be represented
as the region between $V^-_0$ and $V^+_{2 \alpha}$,
where $V^-_0$ and $V^+_{2 \alpha}$ 
are identified by translation.
The function $f(z)$ is
\begin{equation}
f(z) = \frac{2}{\pi} \, \arctan \, z \,,
\end{equation}
and $f \circ \varphi (z)$ denotes
the conformal transformation of $\varphi (z)$ by the map $f(z)$.
The operator ${\cal B}$ is defined by
\begin{equation}
{\cal B} = \int \frac{dz}{2 \pi i} \, b(z) \,,
\end{equation}
and when ${\cal B}$ is located between two operators
at $t_1$ and $t_2$ with $1/2 < t_1 < t_2$,
the contour of the integral can be taken to be $-V^+_\alpha$
with $2 \, t_1 - 1 < \alpha < 2 \, t_2 - 1$.
The anticommutation relation of ${\cal B}$ and $c(z)$ is
\begin{equation}
\{ {\cal B}, c(z) \} = 1 \,,
\end{equation}
and ${\cal B}^2 = 0$.

The state $\Psi^{(n)}$ can be written more compactly as
\begin{equation}
\langle \, \varphi, \Psi^{(n)} \, \rangle
= \int d^{n-1} t \, \Bigl\langle \, f \circ \varphi (0) \,
\prod_{i=0}^{n-2} \Bigl[ \, U (1+ \ell_i) \, {\cal B} \,
\Bigr] \, U (1+ \ell_{n-1}) \,
\Bigr\rangle_{{\cal W}_{1+\ell_{n-1}}} \,,
\end{equation}
where
\begin{equation}
\int d^{n-1} t \equiv
\int_0^1 dt_1 \int_0^1 dt_2 \ldots \int_0^1 dt_{n-1} \,, \quad
\ell_0 = 0 \,, \quad \ell_i \equiv \sum_{k=1}^i t_k \quad
\mbox{for} \quad i=1 \,, \, 2 \,, \, 3 \,, \, \ldots \,.
\label{ell-definition}
\end{equation}
The state $\Psi_\lambda$ can be represented as
\begin{equation}
\Psi_\lambda
= \frac{1}{1- \lambda \, (Q_B X) \, P} \, \lambda \, Q_B X \,,
\end{equation}
where
\begin{equation}
\frac{1}{1- \lambda \, (Q_B X) \, P}
\equiv 1 + \sum_{n=1}^\infty \,
[ \, \lambda \, (Q_B X) \, P \, ]^n \,.
\end{equation}
The state $X$ is described in the CFT language as
\begin{equation}
\langle \, \varphi, X \, \rangle
= \langle \, f \circ \varphi (0) \,\, V (1) \, \rangle_{{\cal W}_1}
= \langle \, f \circ \varphi (0) \,\,
c \, \xi e^{-\phi} V_{1/2} (1) \, \rangle_{{\cal W}_1} \,,
\end{equation}
and the state $Q_B X$ is
\begin{equation}
\langle \, \varphi, Q_B X \, \rangle
= \langle \, f \circ \varphi (0) \,\, Q_B \cdot V (1) \,
\rangle_{{\cal W}_1}
= \langle \, f \circ \varphi (0) \,\,
U (1) \, \rangle_{{\cal W}_1} \,.
\end{equation}
The definition of $P$ is a little involved.\footnote{
The state $P$ corresponds to $J_b$ of \cite{Okawa:2007ri}
in the bosonic case
and to $\eta_0 J$ of \cite{Okawa:2007ri}
in the superstring case.
}
It is defined when it appears as $\varphi_1 \, P \, \varphi_2$
between two states $\varphi_1$ and $\varphi_2$ in the Fock space.
The string product $\varphi_1 \, P \, \varphi_2$ is given by
\begin{equation}
\langle \, \varphi,\, \varphi_1 \, P \,\varphi_2 \, \rangle
= \int_0^1 dt \, \langle \, f \circ \varphi (0) \,
f_1 \circ \varphi_1 (0) \, {\cal B} \, f_{1+t} \circ \varphi_2 (0) \,
\rangle_{{\cal W}_{1+t}} \,,
\end{equation}
where $\varphi_1 (0)$ and $\varphi_2 (0)$
are the operators
corresponding to the states $\varphi_1$ and $\varphi_2$, respectively.
The map $f_a (z)$ is a combination of $f(z)$ and translation:
\begin{equation}
f_a (z) = \frac{2}{\pi} \, \arctan \, z + a \,.
\end{equation}
The string product $\varphi_1 \, P \, \varphi_2$
is well defined if
$f_1 \circ \varphi_1 (0) \, {\cal B} \, f_{1+t} \circ \varphi_2 (0)$
is regular in the limit $t \to 0 \,$.

An important property of $P$ is
\begin{equation}
\varphi_1 \, (Q_B P) \, \varphi_2
= \varphi_1 \, \varphi_2
\label{Q_B-P}
\end{equation}
when $f_1 \circ \varphi_1 (0) \, f_{1+t} \circ \varphi_2 (0)$
vanishes in the limit $t \to 0 \,$.
This relation can be shown in the following way.
Since the BRST transformation of $b (z)$
is the energy-momentum tensor $T(z)$,
the inner product $\langle \, \varphi,\,
\varphi_1 \, (Q_B P) \,\varphi_2 \, \rangle$
is given by
\begin{equation}
\langle \, \varphi,\, \varphi_1 \, (Q_B P) \,\varphi_2 \, \rangle
= \int_0^1 dt \, \langle \, f \circ \varphi (0) \,
f_1 \circ \varphi_1 (0) \, {\cal L} \, f_{1+t} \circ \varphi_2 (0) \,
\rangle_{{\cal W}_{1+t}} \,,
\end{equation}
where
\begin{equation}
{\cal L} = \int \frac{dz}{2 \pi i} \, T(z) \,,
\end{equation}
and the contour of the integral is the same as that of ${\cal B}$.
As discussed in \cite{Okawa:2006vm},
an insertion of ${\cal L}$ is equivalent
to taking a derivative with respect to $t$.
It is analogous to the relation
$L_0 \, e^{-t L_0} = {}-\partial_t \, e^{-t L_0}$
in the standard strip coordinates,
where $L_0$ is the zero mode of the energy-momentum tensor.
We thus have
\begin{equation}
\begin{split}
\langle \, \varphi,\, \varphi_1 \, (Q_B P) \, \varphi_2 \, \rangle
& = \int_0^1 dt \, \partial_t \, \langle \, f \circ \varphi (0) \,
f_1 \circ \varphi_1 (0) \, f_{1+t} \circ \varphi_2 (0) \,
\rangle_{{\cal W}_{1+t}} \\
& = \langle \, f \circ \varphi (0) \,
f_1 \circ \varphi_1 (0) \, f_2 \circ \varphi_2 (0) \,
\rangle_{{\cal W}_2}
= \langle \, \varphi,\, \varphi_1 \, \, \varphi_2 \, \rangle
\end{split}
\end{equation}
when $f_1 \circ \varphi_1 (0) \, f_{1+t} \circ \varphi_2 (0)$
vanishes in the limit $t \to 0 \,$.
This completes the proof of ({\ref{Q_B-P}).
In the language of $\cite{Kiermaier:2007ba}$,
$\varphi_1 \, P \, \varphi_2$ is
\begin{equation}
\varphi_1 \, P \, \varphi_2
= \int_0^1 dt \, \varphi_1 \, e^{-(t-1) L^+_L} \,
(-B^+_L) \, \varphi_2 \,,
\end{equation}
and the relation (\ref{Q_B-P}) follows from
$\{ Q_B ,\, B^+_L \} = L^+_L$.

To summarize, when the regularity conditions
we mentioned are satisfied,
$\Psi_\lambda$ is well defined,
and we can safely use the relation
\begin{equation}
Q_B P = 1
\end{equation}
for the Grassmann-odd state $P$.
It is then straightforward to calculate $Q_B \Psi_\lambda$,
and the result is
\begin{equation}
Q_B \Psi_\lambda
= {}- \frac{1}{1- \lambda \, (Q_B X) \, P} \, \lambda \, (Q_B X) \,
\frac{1}{1- \lambda \, (Q_B X) \, P} \, \lambda \, Q_B X \,.
\end{equation}
We have thus shown that $\Psi_\lambda$
satisfies the equation of motion
for the bosonic string:
\begin{equation}
Q_B \, \Psi_\lambda + \Psi_\lambda^2 = 0 \,.
\end{equation}
Another important property of $\Psi_\lambda$
is that $\eta_0 \Psi_\lambda = 0$.
It is easy to see that $\eta_0$ annihilates
$\Psi^{(n)}$ in (\ref{Psi^(n)})
because $\eta$ and $b$ anticommute
and $U$ is annihilated by $\eta_0$.

\section{Solution}
\setcounter{equation}{0}

Let us now solve
\begin{equation}
Q_B \, G (\lambda) + [ \, \Psi_\lambda , G (\lambda) \, ]
= \Psi'_\lambda \,.
\label{G-equation-2}
\end{equation}
The string field $\Psi'_\lambda$ is given by
\begin{equation}
\Psi'_\lambda \equiv \frac{d}{d \lambda} \, \Psi_\lambda
= \frac{1}{1 - \lambda \, (Q_B X) \, P} \, (Q_B X) \,
\frac{1}{1 - \lambda \, P \, (Q_B X)} \,,
\end{equation}
where
\begin{equation}
\frac{1}{1- \lambda \, (Q_B X) \, P}
\equiv 1 + \sum_{n=1}^\infty \,
[ \, \lambda \, (Q_B X) \, P \, ]^n \,, \quad
\frac{1}{1 - \lambda \, P \, (Q_B X)}
\equiv 1 + \sum_{n=1}^\infty \,
[ \, \lambda \, P \, (Q_B X) \, ]^n \,.
\end{equation}
We look for a solution made of $X$, $P$, and $Q_B$.
We assume for the moment that states involving $P$ are well defined
and that we can use the relation $Q_B P = 1$.
We will discuss regularity conditions
necessary for these assumptions later.
A string field within this ansatz satisfies the reality condition
if it is odd under the conjugation
given by replacing $X \to -X$
and by reversing the order of string products.
Signs from anticommuting Grassmann-odd string fields
have to be taken care of
in reversing the order of string products.
For example, the state $\Psi^{(n)}$ is real
because its conjugation is given by
\begin{equation}
\begin{split}
& \Psi^{(n)} = 
[ \, (Q_B X) \, P \, ]^{n-1} \, Q_B X \\
& \to (-1)^{(2n -1)(n-1)} \,
(- Q_B X) \, [ \, P \, (- Q_B X) \, ]^{n-1}
= {}- [ \, (Q_B X) \, P \, ]^{n-1} \, Q_B X
\end{split}
\end{equation}
for any positive integer $n$.
It is easy to find a perturbative solution
to (\ref{G-equation-2})
by expanding the equation and $G(\lambda)$
in powers of $\lambda$.
We find that the following state
solves (\ref{G-equation-2}) to all orders in $\lambda$
and satisfies the reality condition:
\begin{equation}
G (\lambda) = \frac{1}{1 - \lambda \, (Q_B X) \, P} \, X \,
\frac{1}{1 - \lambda \, P \, (Q_B X)} \,.
\label{G-solution}
\end{equation}
It is easy to see that $G(\lambda)$ in (\ref{G-solution})
solves (\ref{G-equation-2}) from the following relations:
\begin{equation}
\begin{split}
& Q_B \, \frac{1}{1- \lambda \, (Q_B X) \, P}
= {}- \frac{1}{1- \lambda \, (Q_B X) \, P} \,
\lambda \, (Q_B X) \,
\frac{1}{1- \lambda \, (Q_B X) \, P}
= {}- \Psi_\lambda \,
\frac{1}{1- \lambda \, (Q_B X) \, P} \,, \\
& Q_B \, \frac{1}{1 - \lambda \, P \, (Q_B X)}
= \frac{1}{1 - \lambda \, P \, (Q_B X)} \,
\lambda \, (Q_B X) \,
\frac{1}{1 - \lambda \, P \, (Q_B X)}
= \frac{1}{1 - \lambda \, P \, (Q_B X)} \, \Psi_\lambda \,.
\end{split}
\end{equation}
An explicit expression of $G(\lambda)$
in the CFT description is given by
\begin{equation}
\begin{split}
& \langle \, \varphi, G(\lambda) \, \rangle
= \sum_{n=0}^\infty \, \sum_{m=0}^\infty \, \lambda^{n+m} \,
\int d^{n+m} t \, \Bigl\langle \, f \circ \varphi (0) \,
\prod_{i=0}^{n-1} \Bigl[ \, U (1+ \ell_i) \, {\cal B} \,
\Bigr] \, V (1+ \ell_{n}) \\
& \qquad \qquad \qquad \qquad \qquad
\qquad \qquad \qquad \qquad \quad {}\times
\prod_{j=n+1}^{n+m} \Bigl[ \, {\cal B} \, U (1+ \ell_j) \,
\Bigr] \, 
\Bigr\rangle_{{\cal W}_{1+\ell_{n+m}}} \,,
\end{split}
\end{equation}
with the understanding that
\begin{equation}
\prod_{i=0}^{-1} \Bigl[ \, U (1+ \ell_i) \, {\cal B} \,
\Bigr] = 1 \,, \quad
\prod_{j=n+1}^{n} \Bigl[ \, {\cal B} \, U (1+ \ell_j) \,
\Bigr]  = 1 \,, \quad
\int d^0 t = 1 \,.
\end{equation}
Following the strategy outlined in the introduction,
we construct a solution
to the equation of motion (\ref{equation-of-motion})
in open superstring field theory as follows:
\begin{equation}
e^{\Phi_\lambda} = {\rm Pexp} \, \biggl[ \,
\int_0^\lambda d \lambda' \, G(\lambda') \, \biggr] \,,
\end{equation}
or
\begin{equation}
\Phi_\lambda = \ln {\rm Pexp} \, \biggl[ \,
\int_0^\lambda d \lambda' \, G(\lambda') \, \biggr] \,,
\label{solution}
\end{equation}
where our convention for the path-ordered exponential is
\begin{equation}
\begin{split}
& {\rm Pexp} \, \biggl[ \,
\int_a^b d \lambda' \, G(\lambda') \, \biggr]
= 1 + \int_a^b d \lambda_1 \, G(\lambda_1)
+ \int_a^b d \lambda_1 \, \int_a^{\lambda_1} d \lambda_2 \,
G(\lambda_2) \, G(\lambda_1) \\
& \qquad \qquad \qquad \qquad \qquad \quad
{}+ \int_a^b d \lambda_1 \, \int_a^{\lambda_1} d \lambda_2 \,
\int_a^{\lambda_2} d \lambda_3 \,
G(\lambda_3) \, G(\lambda_2) \, G(\lambda_1) \, + \ldots \,.
\end{split}
\label{Pexp-1}
\end{equation}
It can also be written as
\begin{equation}
\begin{split}
& {\rm Pexp} \, \biggl[ \,
\int_a^b d \lambda' \, G(\lambda') \, \biggr]
= 1 + \int_a^b d \lambda_1 \, G(\lambda_1)
+ \int_a^b d \lambda_1 \, \int_{\lambda_1}^b d \lambda_2 \,
G(\lambda_1) \, G(\lambda_2) \\
& \qquad \qquad \qquad \qquad \qquad \quad
{}+ \int_a^b d \lambda_1 \, \int_{\lambda_1}^b d \lambda_2 \,
\int_{\lambda_2}^b d \lambda_3 \,
G(\lambda_1) \, G(\lambda_2) \, G(\lambda_3) \, + \ldots \,.
\end{split}
\label{Pexp-2}
\end{equation}
The path-ordered exponential satisfies
the differential equations given by
\begin{equation}
\begin{split}
& \frac{d}{d b} \, {\rm Pexp} \, \biggl[ \,
\int_a^b d \lambda' \, G(\lambda') \, \biggr]
= {\rm Pexp} \, \biggl[ \,
\int_a^b d \lambda' \, G(\lambda') \, \biggr] \, G(b) \,, \\
& \frac{d}{d a} \, {\rm Pexp} \, \biggl[ \,
\int_a^b d \lambda' \, G(\lambda') \, \biggr]
= {}- G(a) \, {\rm Pexp} \, \biggl[ \,
\int_a^b d \lambda' \, G(\lambda') \, \biggr] \,,
\end{split}
\end{equation}
with the initial condition
\begin{equation}
{\rm Pexp} \, \biggl[ \,
\int_a^b d \lambda' \, G(\lambda') \, \biggr] \,
\biggr|_{a=b} = 1 \,.
\end{equation}
The string field $e^{-\Phi_\lambda}$ is given by
\begin{equation}
e^{-\Phi_\lambda} = {\rm Pexp} \, \biggl[ \,
\int_\lambda^0 d \lambda' \, G(\lambda') \, \biggr] \,.
\end{equation}
It is straightforward to verify that
(\ref{G-equation-2}) can be obtained from
(\ref{Erler-equation})
with $\Phi=\Phi_\lambda$ in (\ref{solution})
by taking a derivative with respect to $\lambda$.
The equation of motion is trivially satisfied
when $\lambda=0$.
Thus $\Phi_\lambda$ in (\ref{solution})
satisfies the equation of motion (\ref{equation-of-motion})
to all orders in $\lambda$. This is the main result of this paper.
We present the expansion of $\Phi_\lambda$ to $O(\lambda^3)$
in appendix A.
While it is guaranteed that $\Phi_\lambda$ satisfies
the reality condition by construction,
we can explicitly confirm this.
Since the conjugate of $G(\lambda)$
associated with the reality condition is $-G(\lambda)$,
the conjugate of $e^{\Phi_\lambda}$
is $e^{-\Phi_\lambda}$, as can be seen using the formulas
(\ref{Pexp-1}) and (\ref{Pexp-2}).
Therefore, its logarithm $\Phi_\lambda$
satisfies the reality condition.

\bigskip

The analytic solution constructed
in \cite{Erler:2007rh, Okawa:2007ri}
can also be written using a path-ordered exponential.
Let us denote the solution in \cite{Erler:2007rh, Okawa:2007ri}
by $\widetilde{\Phi}_\lambda$.
It is given by
\begin{equation}
e^{\widetilde{\Phi}_\lambda} = \frac{1}{1-H_\lambda} \,,
\end{equation}
where
\begin{equation}
H_\lambda =
\frac{1}{1 - \lambda \, (Q_B X) \, P} \, \lambda \, X \,.
\end{equation}
It is easy to calculate $Q_B H_\lambda$ and show that
\begin{equation}
e^{-\widetilde{\Phi}_\lambda} \,
Q_B \, e^{\widetilde{\Phi}_\lambda}
= (Q_B \, H_\lambda) \, \frac{1}{1-H_\lambda} = \Psi_\lambda \,.
\end{equation}
Thus $\widetilde{\Phi}_\lambda$ solves the equation of motion
(\ref{equation-of-motion}). Since
\begin{equation}
\frac{d}{d \lambda} \, e^{\widetilde{\Phi}_\lambda}
= \frac{1}{1-H_\lambda} \, H'_\lambda \, \frac{1}{1-H_\lambda}
= e^{\widetilde{\Phi}_\lambda} \,
H'_\lambda \, \frac{1}{1-H_\lambda} \,,
\end{equation}
where
\begin{equation}
H'_\lambda \equiv \frac{d}{d \lambda} \, H_\lambda \,,
\end{equation}
and $e^{\widetilde{\Phi}_\lambda} = 1$ at $\lambda=0 \,$,
$e^{\widetilde{\Phi}_\lambda}$ can be written as
\begin{equation}
e^{\widetilde{\Phi}_\lambda}
= {\rm Pexp} \, \biggl[ \, \int_0^\lambda d \lambda' \,
\widetilde{G} (\lambda') \, \biggr]
\quad \mbox{with} \quad
\widetilde{G} (\lambda) = H'_\lambda \, \frac{1}{1-H_\lambda} \,.
\end{equation}
It is easy to verify that $\widetilde{G} (\lambda)$ satisfies
(\ref{G-equation-2}) using the following equation:
\begin{equation}
\frac{d}{d \lambda} \, \biggl[ \,
(Q_B \, H_\lambda) \, \frac{1}{1-H_\lambda}
- \Psi_\lambda \, \biggr]
= Q_B \, \biggl( \, H'_\lambda \, \frac{1}{1-H_\lambda} \, \biggr)
+ \biggl[ \, \Psi_\lambda \,,\,
H'_\lambda \, \frac{1}{1-H_\lambda} \, \biggr]
- \Psi'_\lambda= 0 \,.
\end{equation}
We can think of $\Phi_\lambda$ in (\ref{solution})
and $\widetilde{\Phi}_\lambda$ as different choices from
solutions to (\ref{G-equation-2}).

\bigskip

We conclude the section by discussing the regularity conditions
mentioned in the preceding section.
When we proved that $G(\lambda)$ in (\ref{G-solution})
satisfies (\ref{G-equation-2}), we used the following relations:
\begin{equation}
\begin{split}
(Q_B X) \, (Q_B P) \, (Q_B X) & = (Q_B X) \, (Q_B X) \,, \\
(Q_B X) \, (Q_B P) \, X & = (Q_B X) \, X \,, \\
X \, (Q_B P) \, (Q_B X) & = X \, (Q_B X) \,.
\end{split}
\end{equation}
The first two relations were discussed in \cite{Okawa:2007ri},
and they hold if $V_1 (z) \, V_1 (w)$,
$V_1 (z) \, V_{1/2} (w)$, and $V_{1/2} (z) \, V_{1/2} (w)$
are regular in the limit $w \to z$.
The last relation also holds if these conditions are satisfied.
Let us next consider
if the string field $G(\lambda)$ itself is finite
and if any intermediate steps in the proof are well defined.
The expressions can be divergent
when two or more operators collide,
but if the states
\begin{equation}
[ \, ( Q_B X ) \, P \, ]^{n-1} \, (Q_B X) \,, \qquad
[ \, ( Q_B X ) \, P \, ]^{n-1} X \,
[ \, P \, ( Q_B X ) \, ]^{m-1}
\end{equation}
for any positive integers $n$ and $m$ are finite,
the string field $G(\lambda)$ and any intermediate steps
in the proof are well defined.
The conditions for
$[ \, ( Q_B X ) \, P \, ]^{n-1} \, (Q_B X)$
to be finite were discussed in \cite{Okawa:2007ri},
and it is straightforward to extend the discussion
to $[ \, ( Q_B X ) \, P \, ]^{n-1} \, X
[ \, P \, ( Q_B X ) \, ]^{m-1}$.
It is easy to confirm that the $bc$ ghost sector is finite.
For the superghost sector, there is a new term of the form
$\eta e^\phi (1) \, \xi e^{-\phi} (1+\ell_{n-1}) \,
\eta e^\phi (1+\ell_{n+m-2})$, but it is regular as well.
Therefore, all the expressions are well defined
if the contributions from the matter sector listed below
are finite:
\begin{equation}
\begin{split}
& \int_0^1 dt \,
V_{\alpha} (1) \, V_{\gamma} (1+t) \,, \\
& \int d^{n+m} t \,
V_{\alpha} (1) \, \prod_{i=1}^{n-1}
\Bigl[ \, V_1 (1+ \ell_i) \, \Bigr] \,
V_{\beta} (1+ \ell_{n}) \, \prod_{j=n+1}^{n+m-1} \,
\Bigl[ \, V_1 (1+ \ell_j) \, \Bigr] \,
V_{\gamma} (1+ \ell_{n+m})
\end{split}
\end{equation}
for any positive integers $n$ and $m$,
where $V_{\alpha}$, $V_{\beta}$, and $V_{\gamma}$ can be
$V_1$ or $V_{1/2}$,
and we used the notation introduced in (\ref{ell-definition})
with the understanding that
\begin{equation}
\prod_{i=1}^{0} \Bigl[ \, V_1 (1+ \ell_i) \, \Bigr] = 1 \,, \quad
\prod_{j=n+1}^{n} \Bigl[ \, V_1 (1+ \ell_j) \, \Bigr] = 1 \,.
\end{equation}
The only minor difference compared to the conditions
for the solution in \cite{Okawa:2007ri}
is that $V_{1/2}$ can appear three times.
When the string field $G(\lambda)$ is finite,
the solution $\Phi_\lambda$ is also finite
to any finite order in $\lambda$.
We thus conclude that
if operator products of an arbitrary number of $V_1$'s
and at most three $V_{1/2}$'s are regular,
the solution $\Phi_\lambda$ in (\ref{solution})
made of $G(\lambda)$ in (\ref{G-solution})
is well defined and satisfies the equation of motion
(\ref{equation-of-motion}).

\section{Discussion}
\setcounter{equation}{0}

We have constructed analytic solutions for marginal deformations
satisfying the reality condition
in open superstring field theory
when operator products made of $V_1$ and $V_{1/2}$ are regular.
It is important to extend the construction
to the cases where the operator products are singular.
Since the structure of $G(\lambda)$ is
very similar to that of the solutions for the bosonic string
in~\cite{Schnabl:2007az, Kiermaier:2007ba},
we hope that it will not be difficult
to construct solutions for the superstring
once we complete the program
of constructing solutions with singular operator products
developed in~\cite{Kiermaier:2007ba}.\footnote{
It is not clear if the recent approach
to the construction of solutions
with singular operator products in~\cite{Fuchs:2007yy}
can be directly extended to the superstring within our framework.
}

It was important for the approach by Erler \cite{Erler:2007rh}
that the equation of motion in open superstring field theory
(\ref{equation-of-motion}) takes the form
that $\eta_0$ annihilates the pure-gauge configuration
$e^{-\Phi} \, Q_B \, e^\Phi$ of open bosonic string field theory.
Interestingly,
the equation of motion in heterotic string field theory
\cite{Okawa:2004ii, Berkovits:2004xh}
takes the form that $\eta_0$ annihilates a pure-gauge
configuration of closed bosonic string field theory
\cite{Zwiebach:1992ie, Saadi:1989tb, Kugo:1989aa, Kugo:1989tk,
Kaku:1988zv, Kaku:1988zw}.
Therefore, a similar approach may be useful
in constructing solutions in heterotic string field theory
once we find solutions in closed bosonic string field theory.

The open superstring field theory formulated by Berkovits
can also be used to describe the $N=2$ string
by replacing $Q_B$ and $\eta_0$ with the generators
in the $N=2$ string \cite{Berkovits:1995ab}.
The reality condition for the $N=2$ string
is different from that for the ordinary superstring,
and it is not clear if an approach similar to the one
in this paper will be useful
in constructing solutions satisfying the reality condition
for the $N=2$ string.

Open superstring field theory
formulated by Berkovits \cite{Berkovits:1995ab}
is more than ten years old,
and its first analytic solutions have now been constructed.
We expect further exciting developments
in the near future.

\bigskip

\noindent
{\it Note added}

The convention for the conjugation
associated with the reality condition
in this paper and in \cite{Okawa:2007ri}
is different from the one used
in \cite{Erler:2007rh, Kiermaier:2007vu, Kiermaier:2007ki}.
Let us explain the relation
between the two conventions.
The string field must have a definite parity
under the combination of the Hermitean conjugation (hc)
and the inverse BPZ conjugation ($\text{bpz}^{-1}$)
to guarantee that the string field theory action
is real~\cite{Gaberdiel:1997ia}.
If we denote the conjugate of a string field $A$
in this paper and in \cite{Okawa:2007ri} by $A^\ast$,
it is defined by
$$
A^\ast \equiv
\begin{cases}
& \,\,\,\, \text{bpz}^{-1} \circ \text{hc} \, (A) \quad
\text{when the ghost number of}~ A ~\text{is}~
0 ~\text{or}~ 3 ~\text{mod}~ 4 \,, \\
& - \text{bpz}^{-1} \circ \text{hc} \, (A) \quad
\text{when the ghost number of}~ A ~\text{is}~
1 ~\text{or}~ 2 ~\text{mod}~ 4 \,.
\end{cases}
$$
With this definition,
the following relations hold:
$$
(Q_B A)^\ast = Q_B A^\ast \,, \qquad
(A \, B)^\ast = (-1)^{AB} \, B^\ast A^\ast \,,
$$
where $(-1)^{AB} = -1$
when both $A$ and $B$ are Grassmann odd
and $(-1)^{AB} = 1$ in other cases.
If we denote the conjugate of a string field $A$
used in \cite{Erler:2007rh, Kiermaier:2007vu, Kiermaier:2007ki}
by $A^\ddagger$, it is defined by
$$
A^\ddagger \equiv \text{bpz}^{-1} \circ \text{hc} \, (A)
$$
for any ghost number.
With this definition,
the following relations hold:
$$
(Q_B A)^\ddagger = {}- (-1)^A \, Q_B A^\ddagger \,, \qquad
(A \, B)^\ddagger = B^\ddagger \, A^\ddagger \,,
$$
where $(-1)^{A} = -1$
when $A$ is Grassmann odd
and $(-1)^{A} = 1$
when $A$ is Grassmann even.
The open superstring field $\Phi$ has ghost number $0$,
and thus $\Phi^\ddagger = \Phi^\ast$.
The reality condition is satisfied
when $\Phi^\ddagger = \Phi^\ast = -\Phi$.
The open bosonic string field $\Psi$ has ghost number $1$,
and thus $\Psi^\ddagger = - \Psi^\ast$.
The reality condition is satisfied
when $\Psi^\ddagger = - \Psi^\ast = \Psi$.

\bigskip

\noindent
{\bf \large Acknowledgments}

\medskip

I would like to thank Ted Erler for helpful correspondence.

\appendix

\section{Expansion}
\setcounter{equation}{0}

In this appendix we present
the expansion of the solution $\Phi_\lambda$
to third order in $\lambda$.
We first expand $G(\lambda)$ in powers of $X$:
\begin{equation}
\begin{split}
G(\lambda) & = X
+ \lambda \, [ \, (Q_B X) \, P \, X + X \, P \, (Q_B X) \, ] \\
& \quad ~ {}+ \lambda^2 \, [ \, (Q_B X) \, P \, (Q_B X) \, P \, X
+ (Q_B X) \, P \, X \, P \, (Q_B X)
+ X \, P \, (Q_B X) \, P \, (Q_B X) \, ] \\
& \quad ~ {}+ O(X^4) \,.
\end{split}
\end{equation}
The expansion of $e^{\Phi_\lambda}$ is
\begin{equation}
\begin{split}
e^{\Phi_\lambda} & = {\rm Pexp} \, \biggl[ \,
\int_0^\lambda d \lambda' \, G(\lambda') \, \biggr] \\
& = 1 + \int_0^\lambda d \lambda_1 \, G(\lambda_1)
+ \int_0^\lambda d \lambda_1 \, \int_0^{\lambda_1} d \lambda_2 \,
G(\lambda_2) \, G(\lambda_1) \\
& \quad ~ {}+ \int_0^\lambda d \lambda_1 \,
\int_0^{\lambda_1} d \lambda_2 \, \int_0^{\lambda_2} d \lambda_3 \,
G(\lambda_3) \, G(\lambda_2) \, G(\lambda_1)
+ O(X^4) \\
& = 1 + \lambda \, X
+ \frac{1}{2} \, \lambda^2 \,
[ \, (Q_B X) \, P \, X + X \, P \, (Q_B X) + X \, X \, ] \\
& \quad ~ {}+ \lambda^3 \, \biggl[ \,
\frac{1}{3} \, (Q_B X) \, P \, (Q_B X) \, P \, X
+ \frac{1}{3} \, (Q_B X) \, P \, X \, P \, (Q_B X)
+ \frac{1}{3} \, X \, P \, (Q_B X) \, P \, (Q_B X) \\
& \qquad \qquad ~ {}+ \frac{1}{3} \, X \, (Q_B X) \, P \, X
+ \frac{1}{3} \, X \, X \, P \, (Q_B X)
+ \frac{1}{6} \, (Q_B X) \, P \, X \, X
+ \frac{1}{6} \, X \, P \, (Q_B X) \, X \\
& \qquad \qquad ~ {}+ \frac{1}{6} \, X \, X \, X \, \biggr]
+ O(X^4) \,.
\end{split}
\end{equation}
The expansion of the solution $\Phi_\lambda$ is given by
\begin{equation}
\begin{split}
\Phi^{(1)} & = X \,, \\
\Phi^{(2)} & = \frac{1}{2} \,
[ \, (Q_B X) \, P \, X + X \, P \, (Q_B X) \, ] \,, \\
\Phi^{(3)} & = \frac{1}{3} \, (Q_B X) \, P \, (Q_B X) \, P \, X
+ \frac{1}{3} \, (Q_B X) \, P \, X \, P \, (Q_B X)
+ \frac{1}{3} \, X \, P \, (Q_B X) \, P \, (Q_B X) \\
& \quad ~ {}+ \frac{1}{12} \, X \, (Q_B X) \, P \, X
+ \frac{1}{12} \, X \, X \, P \, (Q_B X)
- \frac{1}{12} \, (Q_B X) \, P \, X \, X
- \frac{1}{12} \, X \, P \, (Q_B X) \, X \,.
\end{split}
\label{Phi-expansion}
\end{equation}
Note that $\Phi^{(1)}$, $\Phi^{(2)}$, and $\Phi^{(3)}$
satisfy the reality condition.
The BRST transformation of $\Phi_\lambda$ to $O(\lambda^3)$
is given by
\begin{equation}
\begin{split}
Q_B \, \Phi^{(1)} & = Q_B X \,, \\
Q_B \, \Phi^{(2)} & = (Q_B X) \, P \, (Q_B X)
- \frac{1}{2} \, (Q_B X) \, X
+ \frac{1}{2} \, X \, (Q_B X) \,, \\
Q_B \, \Phi^{(3)} & = (Q_B X) \, P \, (Q_B X) \, P \, (Q_B X) \\
& \quad ~ {}+ \frac{1}{2} \, X \, (Q_B X) \, P \, (Q_B X)
- \frac{1}{2} \, (Q_B X) \, P \, (Q_B X) \, X \\
& \quad ~ {}- \frac{1}{4} \, (Q_B X) \,
[ \, (Q_B X) \, P \, X + X \, P \, (Q_B X) \, ]
+ \frac{1}{4} \,
[ \, (Q_B X) \, P \, X + X \, P \, (Q_B X) \, ] \, (Q_B X) \\
& \quad ~ {}+ \frac{1}{12} \, X \, X \, (Q_B X)
- \frac{1}{6} \, X \, (Q_B X) \, X
+ \frac{1}{12} \, (Q_B X) \, X \, X \,.
\end{split}
\end{equation}
Let us next expand the equation of motion. Since
\begin{equation}
\begin{split}
e^{-\Phi} \, Q_B e^\Phi
& = Q_B \, \Phi
+ \frac{1}{2} \, (Q_B \, \Phi) \, \Phi
- \frac{1}{2} \, \Phi \, (Q_B \, \Phi) \\
& \quad ~ {}+ \frac{1}{6} \, (Q_B \, \Phi) \, \Phi^2
- \frac{1}{3} \, \Phi \, (Q_B \, \Phi) \, \Phi
+ \frac{1}{6} \, \Phi^2 \, (Q_B \, \Phi)
+ O(\Phi^4) \,,
\end{split}
\end{equation}
we have
\begin{equation}
\begin{split}
& \eta_0 \, Q_B \, \Phi^{(1)} = 0 \,, \\
& \eta_0 \, \biggl[ \, Q_B \, \Phi^{(2)}
+ \frac{1}{2} \, (Q_B \, \Phi^{(1)}) \, \Phi^{(1)}
- \frac{1}{2} \, \Phi^{(1)} \, (Q_B \, \Phi^{(1)}) \,
\biggr] = 0 \,, \\
& \eta_0 \, \biggl[ \, Q_B \, \Phi^{(3)}
+ \frac{1}{2} \, (Q_B \, \Phi^{(1)}) \, \Phi^{(2)}
+ \frac{1}{2} \, (Q_B \, \Phi^{(2)}) \, \Phi^{(1)}
- \frac{1}{2} \, \Phi^{(1)} \, (Q_B \, \Phi^{(2)})
- \frac{1}{2} \, \Phi^{(2)} \, (Q_B \, \Phi^{(1)}) \\
& \qquad
{}+ \frac{1}{6} \, (Q_B \, \Phi^{(1)}) \, \Phi^{(1)} \, \Phi^{(1)}
- \frac{1}{3} \, \Phi^{(1)} \, (Q_B \, \Phi^{(1)}) \, \Phi^{(1)}
+ \frac{1}{6} \, \Phi^{(1)} \, \Phi^{(1)} \, (Q_B \, \Phi^{(1)}) \,
\biggr] = 0 \,.
\end{split}
\end{equation}
It is easy to confirm that $\Phi^{(1)}$, $\Phi^{(2)}$,
and $\Phi^{(3)}$ in (\ref{Phi-expansion})
satisfy these equations.

\end{document}